\documentclass[11pt, letterpaper]{article}

\usepackage{amsmath}
\usepackage{amsfonts}
\usepackage{amssymb}
\usepackage{graphicx}

\pdfoutput=1

\usepackage{color}

\usepackage{amsmath, amssymb, graphics, epsfig, graphicx}
\usepackage{epsf}
\usepackage{epstopdf}
\usepackage {amssymb}
\newcommand{\nc}{\newcommand}
\nc{\ba}{\begin{eqnarray}}
\nc{\ea}{\end{eqnarray}}
\newcommand\be{\begin{equation}}
\newcommand\ee{\end{equation}}

\newcommand{\calP}{{\cal{P}}}

\newcommand{\bea}{\begin{eqnarray}}
\newcommand{\eea}{\end{eqnarray}}

\newcommand{\bfq}{{\bf{q}}}

\newcommand{\bfp}{{\bf{p}}}
\newcommand{\Ha}{{\bf H}_3  }
\newcommand{\Hb}{{\bf H}_4  }

\begin{document}

\vspace{5mm}
\vspace{0.5cm}
\begin{center}

\def\thefootnote{\fnsymbol{footnote}}

{ \bf  \large  Revisiting Loop Corrections in Single Field USR Inflation}
\\[1cm]

{ Hassan Firouzjahi$\footnote{firouz@ipm.ir}$}\\[0.5cm]

{\small \textit{ School of Astronomy, Institute for Research in Fundamental Sciences (IPM) \\ P.~O.~Box 19395-5531, Tehran, Iran }}\\

\end{center}

\vspace{.8cm}

\hrule \vspace{0.3cm}


\begin{abstract}

We revisit  the one-loop corrections on CMB scale perturbations induced from small scale modes in single field models which undergo a phase of ultra slow-roll  inflation. There were concerns that large loop corrections are against the notion of the decoupling of  scales and they are  cancelled out once the boundary terms are included in Hamiltonian.   We highlight  that the non-linear coupling between the long and short modes and the modulation of the short mode power spectrum by the long mode are the key physical reasons behind the large loop corrections. In particular, in order for the  modulation by the long mode  to be significant there should be a  strong scale-dependent enhancement in power spectrum of the short mode which is the hallmark of the USR inflation. We  highlight the important roles played by the  would-be decaying mode which were not taken into account properly in recent works claiming the loop cancellation.  
We confirm the original conclusion that the loop corrections are genuine and they can be  dangerous for PBHs formation unless the transition  to the final attractor phase is mild. 

\end{abstract}
\vspace{0.5cm} \hrule
\def\thefootnote{\arabic{footnote}}
\setcounter{footnote}{0}
\newpage
\section{Introduction}
\label{intro}

The question of one-loop corrections in power spectrum in 
models of single field inflation containing an intermediate phase of ultra slow-roll (USR) inflation has attracted considerable interests recently \cite{Kristiano:2022maq, Kristiano:2023scm, Riotto:2023hoz, Riotto:2023gpm, Choudhury:2023vuj,  Choudhury:2023jlt,  Choudhury:2023rks, Choudhury:2023hvf, 
Firouzjahi:2023aum, Motohashi:2023syh, Firouzjahi:2023ahg, Tasinato:2023ukp, Franciolini:2023lgy, Firouzjahi:2023btw, Maity:2023qzw, Cheng:2023ikq, Fumagalli:2023loc, Nassiri-Rad:2023asg, Meng:2022ixx, Cheng:2021lif}. The models incorporating a phase of USR have been employed as a mechanism to enhance the power spectrum on small scales to source the primordial black holes (PBHs) as a candidate for dark matter \cite{Ivanov:1994pa, Garcia-Bellido:2017mdw, Biagetti:2018pjj}, for a review see \cite{Khlopov:2008qy, Ozsoy:2023ryl, Byrnes:2021jka}.

It was argued in \cite{Kristiano:2022maq, Kristiano:2023scm} that the  one-loop corrections induced from small USR modes  can significantly affect the observed CMB scale perturbations. Therefore,  in order to keep these loop corrections under perturbative control, it was argued that the model is not trusted to generate the 
desired PBHs abundance. On the other hand, this conclusion was criticized in  \cite{ Riotto:2023gpm, Riotto:2023hoz} where it was argued that this conclusion 
can not be viewed as a no-go theorem and the dangerous one-loop corrections can be harmless in a smooth transition. This question was studied in further details  in \cite{Firouzjahi:2023aum} in which the effects of both cubic and quartic Hamiltonians were included. In addition, the effects of the sharpness of the transition from the intermediate USR phase to the final attractor phase were highlighted  as well. 
The analysis in \cite{Firouzjahi:2023aum} supports the conclusion of  \cite{Kristiano:2022maq}  when the transition from the USR phase to the final attractor phase is sharp.  However,  it was argued in \cite{Firouzjahi:2023aum} that the 
dangerous one-loop corrections can be washed out  in a mild transition. This question was also studied in  \cite{Firouzjahi:2023ahg} where the one-loop corrections have been calculated   using $\delta N$ formalism. It was shown in \cite{Firouzjahi:2023ahg} that for a mild transition the one-loop corrections are suppressed by the slow-roll parameters so the setup is still reliable 
 for PBHs formations. 
 
On the other hand,  the question of  loop corrections was revisited in \cite{Fumagalli:2023hpa} and \cite{Tada:2023rgp} in which it was claimed that the one-loop correction cancel in the setup of interest. Specifically, in \cite{Fumagalli:2023hpa} the roles of boundary terms were highlighted which were not incorporated in \cite{Kristiano:2022maq, Kristiano:2023scm} and the following works.  On the other hand, in \cite{Tada:2023rgp}, relying on the Maldacena consistency 
condition \cite{Maldacena:2002vr}, it was argued that the large loop  corrections are cancelled once the 
UV limit of the momentum is taken care of by an appropriate $i \varepsilon$ prescription. Note that in both \cite{Fumagalli:2023hpa, Tada:2023rgp}  only the cubic interactions were considered and,  like many other previous works,  the contributions of the quartic interactions are not considered.

In this work  first we study the physical origins of the large loop corrections in this setup. This is more important as the existence of large loop corrections on long CMB scales induced by small scales is somewhat counterintuitive. One may argue that the existence of large loop corrections is against the 
common sense of ``naturalness" and the concept of ``decoupling of scales''. In addition,  we revisit the claims in \cite{Fumagalli:2023hpa, Tada:2023rgp} that the loop corrections cancel and highlight some conceptual and technical points which we  disagree with  these works.  We conclude that the large loop corrections on CMB scales are genuine and  can be dangerous if the transition to the final attractor phase is sharp.


\section{USR Inflation Setup}
\label{setup}

The USR setup is a model of inflation in which the potential is flat \cite{Kinney:2005vj, Morse:2018kda, Lin:2019fcz}. Originally, the USR setup has attracted interests as  a non-trivial  example for the violation of the  Maldacena non-Gaussianity consistency condition \cite{Maldacena:2002vr, Creminelli:2004yq}.  
Since the potential is flat in the USR setup  the inflaton velocity falls off exponentially and the curvature perturbations grow on superhorizon scales \cite{Namjoo:2012aa}. The enhancement of  curvature perturbations  on superhorizon scales is the key behind the violation of the  Maldacena consistency condition  in USR setup \cite{Namjoo:2012aa, Martin:2012pe, Chen:2013aj, Chen:2013eea, Akhshik:2015nfa, Akhshik:2015rwa, Mooij:2015yka, Bravo:2017wyw, Finelli:2017fml, Passaglia:2018ixg, Pi:2022ysn, Firouzjahi:2023xke}.  The amplitude of the local-type non-Gaussianity in USR model is calculated in  \cite{Namjoo:2012aa} to be $f_{NL}=\frac{5}{2}$. This question was further investigated in \cite{Cai:2018dkf} where it was shown that the final amplitude of $f_{NL}$  depends on the sharpness of the transition from the USR phase to the final slow-roll (SR) phase. In particular, in the example of extreme sharp transition from the USR phase to the SR phase, as considered in \cite{Namjoo:2012aa}, $f_{NL}$ acquires its maximum value $\frac{5}{2}$. However, for a mild transition the curvature perturbations evolve after the USR phase until it reaches to its final attractor value. As a result, much of the amplitude of $f_{NL}$ is washed out towards the end of inflation. The important lesson is that  the sharpness of the transition from the USR phase to the final SR phase plays important role when looking at the final amplitude of the cosmological observables. 

The setup we study here, as  in \cite{Kristiano:2022maq, Kristiano:2023scm}, comprises three phases of inflation, $SR \rightarrow USR \rightarrow SR$, 
with a single field inflation driven by the scalar field $\phi$ with the potential 
$V(\phi)$. The first stage of inflation is in the SR phase during which the large CMB scales leave the horizon. This period may take 20-30 e-folds depending on the mass and abundance of PBHs. The curvature perturbation is nearly scale invariant and Gaussian with an amplitude fixed by the COBE normalization. The second stage is the USR phase in which the potential becomes exactly flat and the curvature perturbation grows exponentially to seed the PBHs formation on small scales.  Typically,  the duration of the USR phase is assumed to be about a few e-folds to obtain a sizeable fraction of the dark matter from the PBHs formation.
The USR phase is glued to a second SR phase which is the final stage of inflation.   
Depending on the model parameters, the transition to the final attractor phase can be either mild or sharp which plays important roles for the amplitude of the loop corrections.

Starting with the FLRW metric
\ba
ds^2 = -dt^2 + a(t)^2 d{\bf x}^2 \, ,
\ea
the inflaton field equation in the USR phase is given by
\ba
\ddot \phi(t) + 3 H \dot \phi(t)=0\, , \quad \quad 3 M_P^2 H^2 \simeq V_0, 
\ea
in which $M_P$ is the reduced Planck mass, $H$ is the Hubble  rate during inflation and $V_0$ is the value of the potential during the USR phase. Since $V_0$ is constant, $H$ is nearly constant while 
$\dot \phi \propto a^{-3}$ during the USR phase.  The slow-roll  parameters related to  $H$  are defined as follows,
\ba
\label{ep-eta}
\epsilon \equiv -\frac{\dot H}{H^2} =\frac{\dot \phi^2}{2 M_P^2 H^2}\, , \quad \quad 
\eta \equiv \frac{\dot \epsilon}{H \epsilon} \, . 
\ea
During the  SR phases  both $\epsilon$ and $\eta$ are nearly constant and small.   However, during the USR phase, $\epsilon$ falls off like $a^{-6}$ 
while $\eta\simeq -6$ which is the hallmark of the USR inflation \cite{Kinney:2005vj}. Going to conformal time $d \tau= dt/a(t)$ with $a H \tau \simeq -1$, 
 $\epsilon(\tau)$ is given by as 
 \ba
 \epsilon(\tau) = \epsilon_i \big( \frac{\tau}{\tau_s} \big)^6 \, ,
 \ea
in which $\epsilon_i$ is the value of $\epsilon$  prior to the USR phase. 
We assume the USR phase is extended during the period $\tau_s < \tau <\tau_e$ so  $\epsilon$ at the end of USR phase is $\epsilon_e = \epsilon_i \big( \frac{\tau_e}{\tau_s} \big)^6 $. Defining the number of e-fold as $d N= H dt$, the duration of the USR phase is given by $\Delta N \equiv N(\tau_e) - N(\tau_s)$ so  
$\epsilon_e = \epsilon_i e^{-6 \Delta N}  $.

As in \cite{Cai:2018dkf}, suppose the potential after the USR phase supports a 
period of SR inflation such that 
\ba
V(\phi) = V(\phi_e) + \sqrt{2 \epsilon_V} V(\phi_e)  (\phi -\phi_e) + \frac{\eta_V}{2} V(\phi_e) (\phi -\phi_e)^2 + ... \, .
\ea
Here $2\epsilon_V \equiv  M_P^2\big(V'(\phi_e)/V(\phi_e) \big)^2$ and $\eta_V\equiv  M_P^2 V''(\phi_e)/V(\phi_e)$ are the  usual slow-roll parameters 
defined in terms of the first and second derivatives of the potential. We assume that 
 the potential is continuous at $\phi=\phi_e$. If we further require that the derivative of the potential to be continuous as well then $\epsilon_V=0$ and the transition becomes smooth. However, if $\epsilon_V \neq 0$, then the derivative of the potential is not continuous and there is a kink in the potential. Depending on the value of $\frac{\epsilon_V}{\eta_V}$ the transition can be either mild or sharp. 
 As we are mostly interested in a sharp transition, below we consider $\eta_V=0$. However, this is not a restrictive assumption and most of our analysis will be carried out to the case where $\eta_V\neq0$ as well. 
 
 The background field equation in the final SR phase is given by \cite{Cai:2018dkf}
 (see also \cite{Cai:2022erk})
 \ba
 \frac{d^2 \phi}{ d N^2} + 3 \frac{d \phi}{d N} + 3 M_P \sqrt{2 \epsilon_V} \simeq 0 \, ,
 \quad \quad 3 M_P^2 H^2 \simeq V(\phi_e) \, .
 \ea
Without loss of generality,  assume  the time of the transition to the final SR phase 
to be at $N=0$.  Imposing the continuity of $\phi$ and $\frac{d \phi}{d N}$  at $N=0$, we obtain
\ba
M_P^{-1}\phi(N)= \frac{C_1}{3} e^{-3 N} + \frac{h}{6}  \sqrt{2 \epsilon_V} N + C_2 \, ,
\ea
where the constants of integration $C_1$ and $C_2$ are given by
 \ba
 C_1=  \sqrt{2 \epsilon_e} (1 + \frac{h}{6} ) \, , \quad \quad 
 C_2 = M_P^{-1} \phi_e - \frac{ \sqrt{2 \epsilon_e}}{3} (1 + \frac{h}{6} ) \, .
 \ea
Following \cite{Cai:2018dkf}  we have defined the parameter $h$ as
\ba
\label{h-def}
h\equiv \frac{6 \sqrt{2 \epsilon_V} }{\dot \phi(t_e)} M_P = -6 \sqrt{\frac{\epsilon_V}{\epsilon_e}} \, .
\ea
Since we assume that $\phi$ is decreasing monotonically during inflation, then $\dot \phi<0$ so $h<0$.  As emphasized in \cite{Cai:2018dkf} and \cite{Firouzjahi:2023aum}, $h$ is the key parameter of the setup,  controlling the sharpness of the transition from the USR phase to the final attractor phase. 

The slow-roll parameters, as defined in Eq. (\ref{ep-eta}), in the final SR phase $(N>0)$ are given by
\ba
\label{ep-N}
\epsilon(\tau)= \epsilon_e  \Big(\frac{h}{6} - (1+ \frac{h}{6} ) \big(\frac{\tau}{\tau_e} \big)^3 \Big)^{2} \, ,
\ea
and
\ba
\label{eta-N}
\eta(\tau) = -\frac{6 (6+h)}{(6+h) - h   \big(\frac{\tau_e}{\tau} \big)^3} \, .
\ea
Towards the final stage of inflation, $\tau \rightarrow \tau_0 \rightarrow 0$, we see that $\epsilon \rightarrow \epsilon_e (\frac{h}{6})^2$ while $\eta $ vanishes like $\tau^3$. While $\epsilon$ is  smooth  at the transition point but it is important to note that $\eta$ has a discontinuity  at $\tau=\tau_e$. Just prior to the transition (i.e.  during the USR phase) $\eta=-6$ while right after the transition  $\eta= -6-h$. As a result, near the transition point we can approximate $\eta$ as follows \cite{Cai:2018dkf}
\ba
\eta = -6 - h \theta(\tau -\tau_e) \quad \quad  \tau_e^- < \tau < \tau_e^+ \, .
\ea
Correspondingly, the above approximation yields  
\ba
\label{eta-jump}
\frac{d \eta}{d \tau} = - h \delta (\tau -\tau_e)  \, ,  \quad \quad  \tau_e^- < \tau < \tau_e^+ \, .
\ea

For an infinitely sharp transition $h \rightarrow -\infty$. In this case, $\epsilon$ after the transition evolves rapidly to a larger value so at the end of inflation 
the final value  of $\epsilon$ is given by $\epsilon(\tau_0) \simeq \epsilon_V = \epsilon_e (\frac{h}{6})^2$. 
For an ``instant" sharp transition studied in \cite{Kristiano:2022maq, Kristiano:2023scm}, $h=-6$. 
In this case  $\epsilon$ in the final SR phase is frozen to its value at the end of USR, $\epsilon_e$. 


\section{Origins of Loop Corrections}
\label{origin}

The fact that small scales can induce large loop corrections on long CMB scales is somewhat counterintuitive. Intuitively speaking, based on the concept of ``decoupling of scales", one expects the effects of small scales to be negligible and  under perturbative control. Therefore, it is an important question to ask what are the physical origins of the large loop corrections on long CMB scales? Here we try to answer this question.

\subsection{Non-Linear Long and Short Mode Coupling}
\label{origin}

There are two physical effects as the origins  of the loop corrections on large scales.
The first effect is that there are non-linear couplings between the long and short modes which are inherited from the non-linearity of GR. These non-linear couplings induce source terms for the evolution of the long mode. 
Second, the long mode which leaves the horizon in early stage of inflation  rescales the background expansion so it modulates the power spectrum of the short modes. The combination of these two effects  induce a backreaction on the long mode itself which is the origin of the loop corrections.  This method was nicely employed in  \cite{Riotto:2023hoz}  to calculate the loop corrections which we also follow in this subsection with some modifications.

The cubic action for the curvature perturbation $\zeta$ is given by \cite{Maldacena:2002vr, Kristiano:2023scm}
\ba
S= M_P^2 \int d \tau d^3 x a^2 \epsilon \Big( \zeta'^2 - \big(\partial_i \zeta)^2 + \frac{\eta'}{2} \zeta' \zeta^2 \Big) \, ,
\ea
where here an below, a prime indicates the derivative with respect to the conformal time. Technically speaking, the above action is for $\zeta_n$ defined in \cite{Maldacena:2002vr} which is non-linearly related to $\zeta$ via $\zeta= \zeta_n + {\cal O} (\zeta_n^2)$. The variable $\zeta_n$ is employed to eliminate the boundary terms  \cite{Arroja:2011yj} with the expenses of inducing quartic order Hamiltonians which should be taken care of. Since we look at the cubic interaction at this 
stage to understand the nature of loop corrections, 
the difference is not important and we keep using $\zeta$ instead of 
$\zeta_n$ in this section.   

The evolution of the Fourier space mode function $\zeta_\bfp(\tau)$ to second order in perturbation theory from the above action is given by
\ba
\label{mode-evol}
\zeta_\bfp'' + \frac{( a^2 \epsilon)'}{a^2 \epsilon} \zeta_\bfp' + \frac{( a^2 \epsilon \eta')'}{4 a^2 \epsilon} \int \frac{d^3 \bfq}{( 2 \pi)^3} \zeta_\bfq \zeta_{\bfp- \bfq} 
=0 \, .
\ea 
From the above equation we see a non-linear source term for the evolution of the long mode $\zeta_\bfp$ from the small scale modes 
which plays crucial roles in our discussions below. 

To handle the analysis analytically, we consider a setup with an instant transition from the USR phase to the final attractor phase at $\tau=\tau_e$. In addition, the transition to the final attractor phase is sharp with $|h | \gg 1$. In this limit, there is a delta source in $\eta'$ as given in Eq. (\ref{eta-jump}) so Eq. (\ref{mode-evol}) can be solved easily, yielding \cite{Kristiano:2023scm}
\ba
\label{long-short}
\zeta_{L,\bfp}(\tau_0)= \zeta_{L, \bfp}^{(0)}+ c \int \frac{d^3 \bfq}{(2 \pi)^3} 
\Big[ \zeta^S_{\bfq}(\tau_e)  \zeta^S_{\bfp-\bfq}(\tau_e) - \frac{2}{3 q_e} {\zeta'^S_\bfq}(\tau_e) \zeta^S_{\bfp-\bfq}(\tau_e) \, \Big]  \, ,
\ea
in which $\zeta_L$ and $\zeta^S$ represent the long and short modes, 
$\zeta_L^{(0)}$ represents the linear solution of Eq. (\ref{mode-evol})  in the absence of the mode couplings and $q_e$ represents the scale which leaves the horizon at the end of USR, $q_e= -1/\tau_e$. 
The parameter 
$c$ is a constant which depends on the details of the transition to the final attractor phase which from Eq. (\ref{eta-jump}) is given by  \cite{Kristiano:2023scm} 
$c = -\Delta \eta/4 =-h/4$.  There can be other terms involving either $\zeta_\bfq$ or $\zeta'_\bfq$ at  higher orders in addition to the quadratic source terms given in Eq. (\ref{long-short}). Our analysis can be extended to these general higher order sources as well without limitations.  Finally, we comment that in the analysis of  \cite{Riotto:2023hoz} the last term in Eq. (\ref{long-short}) is further simplified noting that $\zeta'=-(3/\tau)\zeta$ which is valid for the modes which become superhorizon during the USR phase. Here we keep $\zeta'$ since we consider subhorizon modes as well. 

It is important to note that the  left hand side of Eq. (\ref{long-short}) is calculated at the end of inflation $\tau=\tau_0$  while the source terms are calculated at the end of USR phase $\tau=\tau_e$. This is because of our technical assumption that $\eta'$ has a delta source at $\tau=\tau_e$ as  given in Eq. (\ref{eta-jump}).
However, if the transition is not instantaneous and the evolution of $\eta(\tau)$ is continuous, then there are  additional time-integrals when solving for $\zeta_{L,\bfp}(\tau)$ so Eq. (\ref{long-short}) will have a more complicated form \cite{Kristiano:2023scm}. Finally, note that 
there will be additional source terms at the start of USR phase $\tau=\tau_s$ in Eq. (\ref{long-short}) but since the mode function grows only towards the end of USR phase, we can safely ignore the contribution of the source term at $\tau=\tau_s$.

We are interested in two-point functions of the long mode $\zeta_{L,\bfp}$ with 
$\bfp \rightarrow 0$ representing the CMB scale modes. The short modes are denoted by the momentum $\bfq$ which run inside the loop integral with the hierarchy  $p\ll q $.   The power spectrum of the long mode is given by
\ba
\label{power}
\langle {\zeta_{L, \bfp_1}} {\zeta_{L, \bfp_2}} \rangle &=& 
 \langle \zeta_{L, \bfp_1}^{(0)} \zeta_{L, \bfp_2}^{(0)}  \rangle 
+ 2 c   \int \frac{d^3 \bfq}{(2 \pi)^3} \Big[ 
\big\langle \zeta_{L, \bfp_1}^{(0)} \zeta^S_{\bfq}(\tau_e) \zeta^S_{\bfp_2-\bfq}(\tau_e) \big \rangle
-\frac{2}{3 q_e}
\big\langle \zeta_{L, \bfp_1}^{(0)} \zeta'^S_{\bfq}(\tau_e) \zeta^S_{\bfp_2-\bfq}(\tau_e) \big \rangle   \Big] \nonumber\\ &+& {\cal O} (c^2) \, .
\ea
It is understood that  $\zeta^S$ are calculated at $\tau=\tau_e$ while $\zeta_L$ are calculated at $\tau=\tau_0$. The long mode leaves the horizon during the early stage of inflation, long before the USR phase, so $\zeta_L$ is nearly constant. More specifically, the decaying mode actually grows during the USR phase. However, it was suppressed for a long time before the start of USR phase so its enhancement during the short USR phase is not significant enough to compete with the constant mode. As a result we can take $\zeta_L$ to be constant and simply set $\zeta_L(\tau_0) \simeq \zeta_L(\tau_e)$ so all modes in Eq. (\ref{power}) are calculated at $\tau=\tau_e$. As we will see, it is important to realize that there is no time integral in the two-point function (\ref{power}) while the integration is purely over the momentum space.

Finally, with a bit of calculations one can check that the terms in  Eq. (\ref{power}) containing $c^2$ are subleading. A representative contribution of these terms are given by  
\ba
c^2   \int \frac{d^3 \bfq_1}{(2 \pi)^3}  \int  \frac{d^3 \bfq_2}{(2 \pi)^3} \big\langle \zeta^S_{\bfq_1} \zeta^S_{\bfp_1-\bfq_1} \zeta^S_{\bfq_2} \zeta^S_{\bfp_2-\bfq_2} \big \rangle \, ,
\ea
but as they do not contain the extra factor $1/p^3$ we discard these terms.

Now our job is to calculate the three-point correlations 
$\big\langle \zeta_{L, \bfp_1}^{(0)} \zeta^S_{\bfq} \zeta^S_{\bfp_2-\bfq} \big \rangle$ 
and $\big\langle \zeta_{L, \bfp_1}^{(0)} \zeta'^S_{\bfq} \zeta^S_{\bfp_2-\bfq} \big \rangle$
between one long mode and two short modes. To calculate this note that the effects of the long mode is only to rescale the background \cite{Maldacena:2002vr, Creminelli:2004yq}. More specifically, going to comoving gauge, the metric is given by
\ba
ds^2 = - dt^2 + a(t)^2 e^{2 \zeta_L} d{\bf x}^2 \, .
\ea
As we discussed above, the long mode leaves the horizon  long before the USR phase so $\zeta_L$ is nearly constant.  As a result  it can be absorbed into
the space-like coordinate via $x_i \rightarrow e^{\zeta_L} x_i$ so in momentum space $q \rightarrow e^{-\zeta_L} q$. Consequently, the effects of the long mode can be viewed as a modulation for the short modes. More specifically, for the 
three-point correlation $\big\langle \zeta_{L}^{(0)} \zeta^S \zeta^S \big \rangle$ we can write
\ba
\label{long-modulation}
\big\langle \zeta_{L}^{(0)} \zeta^S \zeta^S \big \rangle
\simeq  \big\langle \zeta_{L}^{(0)} \langle  \zeta^S \zeta^S \rangle_{\zeta_L} \big \rangle \simeq \big\langle \zeta_{L}^{(0)} \big \rangle  
 \big \langle \zeta_{L}^{S} \zeta_{L}^{S} \big\rangle + \big\langle \zeta_{L}^{(0)} \zeta_{L}^{(0)} \big \rangle
 \frac{\partial }{\partial \zeta_L} \langle  \zeta^S \zeta^S \rangle \, .
\ea
As $\zeta_{L}^{(0)}$ is statistically incoherent, then $\big\langle \zeta_{L}^{(0)} \big \rangle=0$ and correspondingly the three-point function can be given in terms of the power spectrum $P_\zeta$ as follows, 
\ba
\label{long-modulation2}
\big\langle \zeta_{L}^{(0)} \zeta^S \zeta^S \big \rangle \simeq 
P_{\zeta_L}^{(0)} \frac{\partial   P_{\zeta_S}}{\partial \zeta_L} \, .
\ea

To calculate the other correlations $\big\langle \zeta_{L, \bfp_1}^{(0)} \zeta'^S_{\bfq} \zeta^S_{\bfp_2-\bfq} \big \rangle$ we first symmetrize the non-commutating 
quantum operators $\zeta$ and $\zeta'$ so $2 \zeta' \zeta \rightarrow  \zeta' \zeta +  \zeta \zeta'$, yielding to
\ba
2 \big\langle \zeta_{L}^{(0)} \zeta'^S \zeta^S \big \rangle
= \big\langle \zeta_{L}^{(0)} \zeta'^S \zeta^S \big \rangle + \big\langle \zeta_{L}^{(0)} \zeta^S \zeta'^S \big \rangle \, .
\ea
Following the same logic as above, we obtain 
\ba
\label{long-modulation3}
2 \big\langle \zeta_{L}^{(0)} \zeta'^S \zeta^S \big \rangle = 
P_{\zeta_L}^{(0)}   \frac{\partial  }{\partial \zeta_L}\frac{d P_{\zeta_S}}{d \tau} \, .
\ea

Plugging the relations (\ref{long-modulation2}) and (\ref{long-modulation3})
in our starting equation (\ref{power})  with the understanding that $\bfp_1= -\bfp_2 \rightarrow 0$, we obtain
\ba
\label{power2}
P_{\zeta_L}(\bfp) =  P^{(0)}_{\zeta_L}(\bfp) \Big[ 1 + 
2 c  \int \frac{d^3 \bfq}{(2 \pi)^3}  \Big(   
\frac{\partial  P_{\zeta_S}(\tau_e)}{\partial \zeta_L}  -\frac{1}{3 q_e} 
\frac{\partial  }{\partial \zeta_L}
\frac{d P_{\zeta_S} (\tau_e)}{d \tau}\Big) \Big]  \, .
\ea
As we discussed before, the role of the long mode is to rescale the background quantity so 
\ba
\label{nzeta}
\frac{\partial  P_{\zeta_S}}{\partial \zeta_L}  = -\frac{\partial  P_{\zeta_S}}{\partial \ln q}   = \big( 1 -n_{\zeta} \big) P_{\zeta_S} \, ,
\ea
in which $n_\zeta$ represents the scale-dependence of the short modes.

Plugging Eq. (\ref{nzeta}) in Eq. (\ref{power2}), and defining 
is the dimensionless power spectrum $\calP_\zeta$ related to $P_\zeta$ as,
\ba
\calP_\zeta(q) \equiv \frac{q^3}{2 \pi^2} P_\zeta(q) \, ,
\ea
we obtain 
\ba
\label{power3}
\calP_{\zeta_L}(\bfp) =  \calP^{(0)}_{\zeta_L}(\bfp) \Big[ 1 - 2 c \int  d \ln q \, 
 \Big(  \frac{\partial   \calP_{\zeta_S} }{\partial \ln q} 
-\frac{1}{3 q_e} 
\frac{\partial }{\partial \ln q} \frac{d  \calP_{\zeta_S}}{d \tau} \Big)\Big|_{\tau=\tau_e}  \Big]  \, .
\ea
The integral above is in the form of a  total derivative, yielding to the following fractional loop correction  in long mode power spectrum
\ba
\label{loop}
\frac{\Delta \calP_{\zeta_L}}{\calP_{\zeta_L}} = -2 c \int \Big( d \calP_{\zeta_S}-\frac{1}{3 q_e}d \calP'_{\zeta_S}\Big)\Big|_{\tau=\tau_e} \, .
\ea
Defining the ``modified" power spectrum $\overline{\calP}$ via  
\ba
\label{barP}
 \overline{\calP}_{\zeta_S}(q, \tau_e) \equiv 
 \calP_{\zeta_S}(q, \tau_e) - \frac{1}{3 q_e}  \calP'_{\zeta_S}(q, \tau_e) \, , 
\ea
the loop correction  in long mode power spectrum is given by
\ba
\label{loop-b}
\frac{\Delta \calP_{\zeta_L}}{\calP_{\zeta_L}} 
= - 2 c \Big[  \overline{\calP}_{\zeta_S}(q_{\mathrm{max}}, \tau_e) -  \overline{\calP}_{\zeta_S}(q_{\mathrm{min}}, \tau_e) \Big]  \, ,
\ea
in which $q_{\mathrm{max}}$ and $q_{\mathrm{min}}$ represent the higher UV and the lower IR regimes of the integration over the short modes. A similar result was originally obtained in \cite{Riotto:2023hoz} who simplified the second term in 
Eq. (\ref{long-short}) via $\zeta'=-(3/\tau)\zeta$ which is valid for the modes which become superhorizon during the USR phase. Here, since we need to consider the subhorizon modes as well, we keep the contribution of $ \calP'_{\zeta_S}(q, \tau_e)$ in $\overline{\calP}_{\zeta_S}(q, \tau_e)$ in its general form.  Having said this, we note that for practical purposes $\overline{\calP}_{\zeta_S}(q, \tau_e) \sim {\calP}_{\zeta_S}(q, \tau_e)$. The above result is also in line with the result obtained in \cite{Tada:2023rgp}  who assumed  Maldacena's consistency condition for the original field $\zeta$. From  Eq. (\ref{power3}) we see that in order for the loop correction to be significant, we require a strong scale-dependent for the short modes. In other words, only the small scales which show scale-dependence will contribute to the integral in Eq. (\ref{power3}). 

In estimating the loop corrections, as in \cite{Kristiano:2022maq, Kristiano:2023scm}, a good prescription is to consider the modes which become superhorizon during the USR phase, corresponding to $q_{\mathrm{max}}= q_e=-1/\tau_e$ and $q_{\mathrm{min}} = q_s=-1/\tau_s$. If we use this prescription, and noting that the power increases exponentially during the USR phase so $\overline{\calP}_\zeta(q_e, \tau_e)  \sim \calP(q_e, \tau_e) \sim e^{6 \Delta N} \calP_\zeta(q_s) $, then one can safely ignore the contribution from the lower bound of the integral $(q_{\mathrm{min}})$ and
\ba
\label{loop2}
\frac{\Delta \calP_{\zeta_L}}{\calP_{\zeta_L}}  \sim -2 c \calP_\zeta(q_e, \tau_e)  \sim e^{6 \Delta N} \calP_{\mathrm{CMB}} \, ,
\ea
in which $\calP_{\mathrm{CMB}} \sim 2\times 10^{-9}$ is the power spectrum on the CMB scales. 
The above result is qualitatively in agreement with the results of \cite{Kristiano:2022maq, Kristiano:2023scm}, highlighting the dangerous one-loop correction if one 
considers  large enough  value of $\Delta N$, i.e. a long enough period of USR inflation, so the factor $ e^{6 \Delta N} \calP_{\mathrm{CMB}}$ can become order unity.

The authors of \cite{Tada:2023rgp} argued that the contribution of the UV part in Eq. (\ref{loop}) is negligible after implementing the usual $i \varepsilon$ prescription 
$\tau \rightarrow (1+ i \varepsilon \tau)$ such that  $e^{- i q \tau} \rightarrow e^{-i q\tau + \varepsilon q \tau}$  so the UV contribution becomes negligible. However, 
by Looking at our derivation of the loop correction in Eq. (\ref{loop-b}), this prescription is  unjustified as we have no integration over $\tau$. 
More specifically, the $i \varepsilon$ prescription $\tau \rightarrow (1+ i \varepsilon \tau)$ is usually performed to kill the 
rapid oscillations in the UV region when one is dealing with an integral over $\tau$. However, in our analysis,  there is no integration over $\tau$. This is because the source term in  Eq. (\ref{mode-evol}) receives a delta source at 
$\tau=\tau_e$ so all  the mode functions are calculated at a fixed time $\tau=\tau_e$. The rapid oscillations occur only in $q$ space since modes which are deep inside the horizon at the time $\tau_e$ still experience the Minkowski background and naturally they oscillate rapidly. As we demonstrate shortly, there will be a quadratic divergence in the momentum space which should be regularized as in standard QFT analysis. In essence this is similar to regularizing the quartic divergence associated to short modes when dealing with the vacuum zero point energy and the cosmological constant problem.

Motivated by discussions in \cite{Tada:2023rgp} now suppose we do not follow the prescription of \cite{Kristiano:2022maq, Kristiano:2023scm} and  take $q_{\mathrm{max}}$ to the extreme UV value allowed. In dimensional regularization approach, $q_{\mathrm{max}}$ can go to infinity. However, in a simple regularization  employing  UV momentum cut off approach,  the largest allowed value of $q_{\mathrm{max}}$ is $q_f$, the mode which leaves the horizon just at the end of inflation, $\tau= \tau_0\rightarrow 0$. Then, the question is what is the power spectrum for that scale  at $\tau=\tau_e$ i.e. $\overline{\calP}_\zeta(q_f, \tau_e)$?   To answer this question, we have to calculate the  mode function $\zeta_q (\tau)$ for the small scale modes, i.e. modes which are  subhorizon during the USR phase 
but become superhorizon  after the USR phase.

\subsection{Mode Function after the USR Phase}
\label{mode}

To obtain the  outgoing mode function, we have to impose the matching conditions at the start and at the end of USR phase for an arbitrary mode $q$.

Starting with the Bunch-Davies initial condition during the first phase of inflation, 
the mode function in the USR phase is given by \cite{Firouzjahi:2023aum}
\ba
\label{mode2}
\zeta^{(2)}_{q} =  \frac{H}{ M_P\sqrt{4 \epsilon_i q^3}}  \big( \frac{\tau_s}{\tau} \big)^3
\Big[ \alpha^{(2)}_q ( 1+ i q \tau) e^{- i q \tau}  + \beta^{(2)}_q ( 1- i q \tau) e^{ i q \tau}  \Big]  \, ,
\ea
with the coefficients $\alpha^{(2)}_q$ and $\beta^{(2)}_q$ given by 
\ba
\label{alpha-beta2}
\alpha^{(2)}_q = 1 + \frac{3 i }{ 2 q^3 \tau_s^3} ( 1 + q^2 \tau_s^2) \, , \quad \quad
\beta^{(2)}_q= -\frac{3i }{ 2 q^3 \tau_s^3 } {( 1+ i q \tau_s)^2} e^{- 2 i q \tau_s} \, .
\ea 
The mode function after the USR phase is given by 
\ba
\label{mode3}
\zeta^{(3)}_{q} =  \frac{H}{ M_P\sqrt{4 \epsilon(\tau) q^3}}  
\Big[ \alpha^{(3)}_q ( 1+ i q \tau) e^{- i q \tau}  + \beta^{(3)}_q ( 1- i k \tau) e^{ i q \tau}  \Big] \, ,
\ea
in which $\epsilon(\tau)$ is given by Eq. (\ref{ep-N}) while $\alpha^{(3)}_q$ and $\beta^{(3)}_q$ are given by,
\ba
\label{alpha-beta3}
\alpha^{(3)}_q = \frac{1}{8 q^6 \tau_s^3 \tau_e^3}  \Big[ 3h
 ( 1 -i q \tau_e)^2 (1+i q \tau_s)^2 e^{2i q (\tau_e- \tau_s)}
-i (2 q^3 \tau_s^3 + 3i q^2 \tau_s^2 + 3 i) (4 i q^3 \tau_e^3- h q^2 \tau_e^2 - h) \Big]
\nonumber
\ea
and
\ba
\beta^{(3)}_q=   \frac{-1}{8 q^6 \tau_s^3 \tau_e^3}  \Big[ 3 ( 1+ i q \tau_s)^2 ( h+ h q^2 \tau_e^2 + 4 i q^3 \tau_e^3 ) e^{-2 i q \tau_s} + i h ( 1+ i q \tau_e)^2  ( 3 i + 3 i q^2 \tau_s^2 + 2 q^3 \tau_s^3 ) e^{- 2 i q \tau_e}
 \Big] \nonumber
\ea 
To calculate the loop corrections in long mode power spectrum from Eq. (\ref{loop-b}), we only need $\zeta_q^{(2)}(\tau_e)$ to calculate $\overline{\calP}_\zeta(q, \tau_e)$. However, for later purposes, we also calculate the outgoing power spectrum at the end of inflation $\tau= \tau_0\rightarrow 0$,  given by
\ba
\label{power-final}
\calP_\zeta(q, \tau_0) = \frac{H^2}{ 8 M_P^2 \pi^2  \epsilon_\mathrm V }  
 \big| \alpha^{(3)}_q   + \beta^{(3)}_q \big|^2  \, ,
\ea
in which $\epsilon_V$ is the value of the slow-roll parameter at the end of inflation.

\begin{figure}[t]
\vspace{-0.5 cm}
	\centering
	\includegraphics[ width=0.85\linewidth , height=0.69\linewidth]{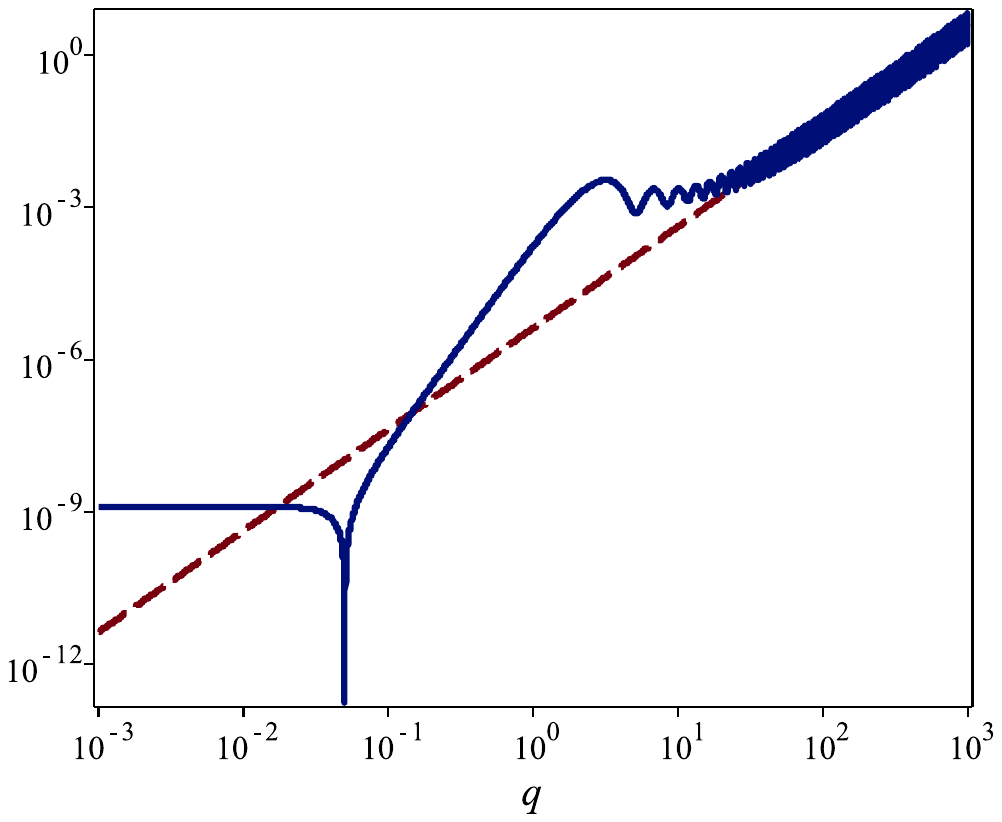}
	\vspace{-4.5 cm}
	\caption{   The (log-log) plot of the modified power spectrum $\overline{\calP}_\zeta(q, \tau_e)$ calculated at $\tau_e$ for $h=-6$ and $\Delta N=\ln(10) \simeq 2.3$. The USR phase starts at $q=1$ while the small scale modes which leave the horizon long after the USR phase correspond to $q \gg 1$.   The red dashed line represents the overall factor in Eq. (\ref{app-1}) indicating the $q^2$ divergence for the UV modes while the rapid oscillations superimposed on top of this scaling can be sen as well.   The position of the dip is prior to the USR phase and the rapid rise of the power spectrum $\calP_\zeta \propto q^4$ prior to the peak is the hallmark of the USR setup.   
}
\label{fig}
\end{figure}

With the mode function $\zeta_q^{(2)}(\tau_e)$ at hand, we can calculate the one-loop corrections from Eq. (\ref{loop-b}). A schematic plot of $\overline{\calP}(q, \tau_e)$
is presented in Fig. \ref{fig}. For the modes which leave the horizon during the first stage of inflation $q \ll q_s$, the power spectrum has a plateau given by the COBE normalization $\calP_{\mathrm{CMB}}$. There is a dip prior to the USR phase and a sharp rise in the power spectrum in the intermediate USR phase, followed by a peak  with oscillations superimposed. All these properties are well-understood, see for example \cite{Byrnes:2018txb, Cole:2022xqc, Carrilho:2019oqg, Ozsoy:2021pws, Pi:2022zxs}. In particular, the power spectrum grows like $\calP_\zeta \propto q^4$ just prior to the peak. This is essential for the loop corrections as given in Eq. (\ref{nzeta}). On the other hand, the modes with $q > q_e$ are subhorizon during the USR phase and the power spectrum grows like $q^2$ with rapid oscillations superimposed on top of it. More specifically, for $q \gg q_e$, we have
\ba
\label{app-1}
| \overline{\calP}_\zeta(q, \tau_e)| \simeq \frac{1}{3} \big( \frac{q}{q_e} \big)^2  e^{6 \Delta N}\calP_{\mathrm{CMB}} \Big[ 1- 6 \cos\big( \frac{2q}{q_s} \big) e^{-3 \Delta N}
\Big] \, ,
\ea
so we see a quadratic divergence in loop corrections for $q \rightarrow \infty$  while 
the rapid oscillations have subleading amplitudes. These behaviours can be seen in 
in Fig. \ref{fig} as well.  

The quadratic divergence of power spectrum for the UV scale is expected 
which is the hallmark of the QFT corrections. To find a finite physical result, we have to renormalize the divergent loop corrections order by order. However, in order for the renormalization procedure to work at each order, we have to make sure that the starting one-loop corrections is under control. In order for the one-loop corrections to be small we require $e^{6 \Delta N}\calP_{\mathrm{CMB}}\ll 1$ since this quantity controls the common amplitude  of the loop 
corrections as can be seen in both Eqs. (\ref{loop2}) and (\ref{app-1}).  
To perform the renormalization, one may set $q_{\mathrm{max}}= q_f$, the modes which leave the horizon at the end of inflation. In this way, one counts the contribution of all modes which become superhorizon by the end of inflation.

It is important to note that the final renormalized loop corrections is not necessarily zero.  On the other hand, it was argued in \cite{Tada:2023rgp} that the leading loop corrections vanish after one kills the rapid oscillations by an $i \varepsilon$ prescription on $\tau$. As we argued previously, this is unjustified. First, we have no integration over the time coordinate as the mode functions in 
Eq. (\ref{loop-b}) are calculated at a fixed time $\tau=\tau_e$. 
Second, the rapid oscillations in $q$ are subleading compared to the dominant quadratic divergence so the renormalized power spectrum is not zero. Finally, one has to employ the standard QFT methods, such as the dimensional regularization scheme, to regularize and renormalize the quadratic divergence. For earlier works concerning the loop corrections and renormalizations in slow-roll setup see 
 \cite{Weinberg:2005vy, Senatore:2009cf, Pimentel:2012tw}. 

\begin{figure}[t]
\vspace{-0.5 cm}
	\centering
	\includegraphics[ width=0.85\linewidth , height=0.69\linewidth]{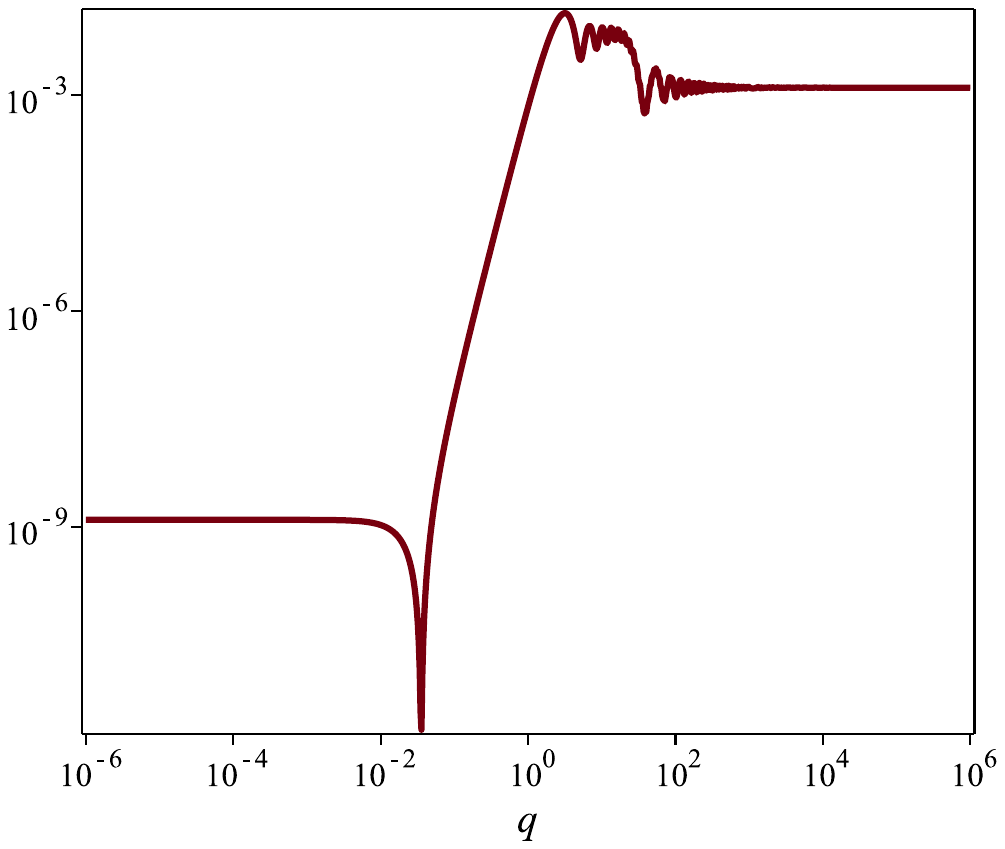}
	\vspace{-4.5 cm}
	\caption{   Power spectrum $\calP_\zeta(q, \tau_0)$ measured at the time of end of inflation $\tau=\tau_0$, for $h=-6$ and $\Delta N=\ln(10) \simeq 2.3$ with USR phase starting at $q=1$. For modes leaving the horizon by the end of USR phase with 
	$q \lesssim 1$ the behaviour is the same as in Fig. \ref{fig} while for small scales with $q\gg 1$ there is a significance difference where the 
	power spectrum reaches an asymptotic value given by Eq. (\ref{power-final2}).
}
\label{fig2}
\end{figure}

While the power spectrum at $\tau=\tau_e$ has a behaviour as shown in Fig. \ref{fig}, but it is also instructive to look at the final power spectrum $\calP_\zeta(q, \tau_0)$ 
measured at the time of end of inflation, $\tau=\tau_0$. A schematic view of the power spectrum is  presented in Fig. \ref{fig2}. 
We see that for modes which leave the horizon by the end of USR phase with 
$q \lesssim q_e$ the power spectrum is similar to Fig. \ref{fig}. However, for the modes which become superhorizon after the USR phase the power spectrum shows a significant difference in which it reaches a plateau instead of growing quadratically. 
Defining $x\equiv - q \tau_s$, the final power spectrum for  $ x \gg1$  is given by 
\ba
\label{power-final2}
\calP_\zeta(q, \tau_0) &\simeq&  {e^{6 \Delta N}} {\cal P}_{CMB} \big(\frac{h-6}{h}\big)^2 \Big(1 + 3 \frac{ \sin(2 x)}{x} \Big)  \nonumber\\
&\simeq& \frac{H^2}{8 \pi^2 \epsilon_V M_P^2 } \big(\frac{h-6}{6}\big)^2 \, .
\ea
The power spectrum reaches a plateau given by Eq. (\ref{power-final2}). In addition, 
 the sharper is the transition, the larger is the final value of the power spectrum \cite{Cai:2018dkf}. 
 
In conclusion, we reproduce the results in \cite{Kristiano:2022maq, Kristiano:2023scm} indicating that the loop correction is a genuine phenomena. The loop corrections are under perturbative control if $e^{6 \Delta N}\calP_{\mathrm{CMB}}\ll 1$.

After performing the technical analysis, here we summarize the physical reasons behind the non-trivial loop correction. There are two important effects which should be taken into account: the non-linear coupling between the long and short modes which provide the source term for the evolution of the long mode. Second,  the long mode provides a modulation to the spectrum of the short mode. This modulation becomes significant if the power spectrum of the short mode experiences a significant scale-dependent enhancement. In our case at hand, this corresponds to a maximum scale-dependent $\calP_\zeta \propto q^4$ just prior to the peak of the power spectrum. Finally, the combination of the non-linear coupling between the 
long and short modes and the modulation of the short modes by the long mode back-reacts on the long mode itself and induces the  one-loop correction. This picture was first put forward in \cite{Riotto:2023hoz}, see also  \cite{Franciolini:2023lgy}.

\section{One-Loop Correction from Cubic Hamiltonian}
\label{one-loop}
 
In this section we revisit the analysis of \cite{Fumagalli:2023hpa} who calculated the 
one-loop corrections from the cubic interaction Hamiltonian  and concluded that the loop corrections cancel out. 
 
\subsection{In-In Analysis}
\label{in-in}

 To calculate the loop corrections, we employ the standard in-in formalism \cite{Weinberg:2005vy} in which the expectation value of the operator $\hat {O}$ at the end of inflation $\tau_0$ is given by the following perturbative  series,
 \ba
 \label{Dyson}
 \langle \hat O(\tau_0) \rangle = \Big \langle \Big[ \bar {\mathrm{T}} \exp \Big( i \int_{-\infty}^{\tau_0} d \tau' H_{in} (\tau') \Big) \Big] \,  \hat O(\tau_0)  \, \Big[ \mathrm{T} \exp \Big( -i \int_{-\infty}^{\tau_0} d \tau' H_{in} (\tau') \Big) \Big] 
 \Big \rangle \, ,
 \ea
 in which $\mathrm{T}$ and $\bar {\mathrm{T}}$ represents the time ordering and anti-time ordering respectively while $H_{in}(\tau)$ represents the interaction Hamiltonian which in our case is $H_{in}(\tau) = {\bf H}_3 $. 
 
The one-loop correction from the cubic self-interaction is calculated in \cite{Fumagalli:2023hpa} in two different methods.
The first method incorporates the boundary terms directly in the cubic Hamiltonian, yielding to
\ba
\label{H3-1}
 {\bf H}_3  = H_{a} + H_b
\ea
in which $H_a$ is a bulk term given by 
\ba
\label{Ha}
H_a= -M_P^2 \int d^3 x  \big( \frac{a^2 \epsilon} {2} \eta' \zeta^2 \zeta' \big) \, ,
\ea
while $H_b$ is a boundary term \cite{Arroja:2011yj}:   
\ba
\label{Hb}
H_b= M_P^2 \int d^3 x   \frac{d}{d \tau} \Big( \frac{a^2 \epsilon} {2} \eta \zeta^2 \zeta' \Big)  \, .
\ea
Note that here, following \cite{Fumagalli:2023hpa}, we work with $\zeta$ itself while in previous sections we were working with $\zeta_n$.  
The effects of the boundary term  $H_b$ were not considered in  \cite{Kristiano:2022maq, Kristiano:2023scm}. It was argued in \cite{Fumagalli:2023hpa} that
once its contributions are added along with the bulk term $H_a$, then the one-loop corrections in power spectrum cancel each other to order $p^3/q^3 \ll1$.

In the second method used in \cite{Fumagalli:2023hpa}, the term containing $\eta'$ in $H_a$ is traded via a boundary term. The new boundary term cancels exactly the boundary term $H_b$. After using the linear field equation,  one obtains the  following equivalent Hamiltonian:  
\ba
\label{H3-2}
 {\bf H}_3  = H_{c} + H_d =  M_P^2 \int d^3 x \,  {a^2 \epsilon}  \eta\, 
 \Big( \zeta'^2 \zeta  + \frac{1}{2} \zeta^2  \partial^2\zeta  \Big)  \, ,
\ea
in which  $H_c$ and $H_d$ are both bulk terms, given by 
\ba
\label{Hc}
H_c= M_P^2 \int d^3 x  {a^2 \epsilon}  \eta\,  \zeta'^2 \zeta  \, ,
\ea
and  
\ba
\label{Hb}
H_d= \frac{M_P^2}{2} \int d^3 x  {a^2 \epsilon}  \eta \zeta^2  \partial^2\zeta  \, .
\ea
Interestingly, we see that the Hamiltonian (\ref{H3-2}) is exactly the same as the cubic Hamiltonian obtained in \cite{Akhshik:2015nfa, Firouzjahi:2023aum} in which $\zeta = - H \pi + {\cal O}(\pi^2) $ where 
$\pi$ represents the Goldstone boson associated to the fluctuations of the inflaton field. On the other hand, no cancellation of the one-loop corrections were observed in  \cite{Firouzjahi:2023aum} at the cubic order (and even in the presence of quartic interactions). This rings the bell that something is going wrong either in the analysis of  \cite{Firouzjahi:2023aum} or \cite{Fumagalli:2023hpa}. Here, we pay careful attention to find the source of disagreement between \cite{Firouzjahi:2023aum} and \cite{Fumagalli:2023hpa} and whether or not the one-loop corrections cancel out
as claimed  in \cite{Fumagalli:2023hpa}.

To perform the in-in analysis, the following relations for $\tau_s \leq \tau \leq \tau_e$
have been used in 
\cite{Fumagalli:2023hpa}:
\ba
\label{com1}
\big [\zeta_\bfq(\tau), \zeta'_\bfp(\tau_0) \big]  = (2 \pi)^3 \delta^3( \bfq+ \bfp) \frac{i}{2 a^2 M_P^2 \epsilon(\tau)} \, ,
\ea
and
\ba
\label{com2}
\big [\zeta_\bfq(\tau), \zeta_\bfp(\tau_0) \big]  \simeq 0 \, .
\ea
A careful investigation shows that Eq. (\ref{com1}) is correct but Eq. (\ref{com2}) is incorrect. Indeed, it was argued in \cite{Fumagalli:2023hpa} that since for the long mode $\zeta_\bfp$ is nearly conserved on superhorizon scales, then $\zeta_\bfp(\tau_0)  \simeq \zeta_\bfp(\tau)$ for $\tau_s \leq \tau \leq \tau_e$, and since the equal time  commutator of the field vanishes, then one obtains Eq. (\ref{com2}). However, there is a subtle flaw in this argument in which the roles of the decaying mode is neglected during the USR phase. More specifically, the decaying mode will grow during the USR phase so  the approximation $\zeta_\bfp(\tau_0) \simeq \zeta_\bfp(\tau)$ may not be consistent  when one is dealing with a nested integral. As shown in   \cite{Firouzjahi:2023aum} (see Eq. (A.23) in  \cite{Firouzjahi:2023aum}), for $\tau_s \leq \tau \leq \tau_e$ one instead has
\ba
\label{com2b}
\big [\zeta_\bfq(\tau), \zeta_\bfp(\tau_0) \big]  \simeq (2 \pi)^3 \delta^3( \bfq+ \bfp)
\frac{i  \tau}{6 a^2 M_P^2  \epsilon(\tau) } \Big( 1 + \frac{6-h}{h} \frac{\tau^3}{\tau_e^3} \Big) \, .
\ea
Comparing  Eq. (\ref{com2b}) with Eq. (\ref{com1}), it is not guaranteed that inside the nested integral, one can automatically neglect Eq. (\ref{com2b}) while keeping Eq. (\ref{com1}). 

In the following, we repeat the analysis of \cite{Fumagalli:2023hpa} 
using the Hamiltonian given in Eq. (\ref{H3-2}). We also compare the results with those in  \cite{Firouzjahi:2023aum} which were obtained via a different in-in method.

 The analysis in \cite{Fumagalli:2023hpa} is based on the commutator approach \cite{Weinberg:2005vy} in which 
 \ba
\big \langle \zeta_\bfp(\tau_0)  \zeta_{\bfp'} (\tau_0)  \big\rangle =  -\int^{\tau_0} d \tau_1 \int^{\tau_1} d \tau_2  \Big \langle \Big[ {\bf H_3} (\tau_2) \,  , \Big[  {\bf H_3} (\tau_1) \, , 
\zeta_\bfp(\tau_0)  \zeta_{\bfp'} (\tau_0)   \Big]\,  \Big ] \Big \rangle  \, .
\ea

Depending on where $H_c$ and $H_d$ are located in the nested integrals,
we obtain
\ba
\label{loop-terms}
\big \langle \zeta_\bfp  \zeta_{\bfp'}  \big\rangle =  
\big \langle \zeta_\bfp  \zeta_{\bfp'}  \big\rangle_{[c,c]} +  
\big \langle \zeta_\bfp  \zeta_{\bfp'}  \big\rangle_{[d,c]} + 
\big \langle \zeta_\bfp  \zeta_{\bfp'}  \big\rangle_{[c,d]}
\ea
in which, for example, 
\ba
\big \langle \zeta_\bfp  \zeta_{\bfp'}  \big\rangle_{[c,c]} = 
 -\int^{\tau_0} d \tau_1 \int^{\tau_1} d \tau_2  \Big \langle \Big[ {\bf H_c} (\tau_2) \,  , \Big[  {\bf H_c} (\tau_1) \, , 
\zeta_\bfp(\tau_0)  \zeta_{\bfp'} (\tau_0)   \Big]\,  \Big ] \Big \rangle  \, ,
\ea
and so on.

In the analysis of \cite{Fumagalli:2023hpa} it was argued that the first two 
terms in  Eq. (\ref{loop-terms}) cancel each other to the volume order $p^3/q^3\ll1$
while the last term in  Eq. (\ref{loop-terms}) is subleading. Indeed, as we check specifically below,  the conclusion that $\big \langle \zeta_\bfp  \zeta_{\bfp'}  \big\rangle_{[c,d]}$ is subleading is correct. However, we show that  the cancellation between the first two terms in Eq. (\ref{loop-terms}) is not exact which is the source of the discrepancy between the result of \cite{Firouzjahi:2023aum} and \cite{Fumagalli:2023hpa}.

To perform the in-in analysis, as in \cite{Firouzjahi:2023aum} and \cite{Fumagalli:2023hpa},  we only consider the contributions for the time interval 
$\tau_s \leq \tau \leq \tau_e$. Performing all the contractions and incorporating the symmetric factors, one obtains
\ba
\label{loop-cc}
\big \langle \zeta_\bfp  \zeta_{\bfp'}  \big\rangle'_{[c,c]} = 
- 8  M_P^4  \int_{\tau_s}^{\tau_e} d \tau_1 \int_{\tau_s}^{\tau_1} d \tau_2  \int \frac{d^3 \bfq}{(2 \pi)^3} \mathrm{Im} \Big[ X_1^* ( \tau_2) \Big(c_1 \, \delta(\tau_1) + 2  \beta(\tau_1)   \Big) \Big]  \, ,
\ea
in which the coefficient $c_1$ is added for bookkeeping as we discuss below. Here and below, $\langle ... \rangle'$ means we absorbed the overall 
factor $(2 \pi)^3 \delta^3( \bfq+ \bfp)$.  In addition, 
\begin{equation}
\label{X1-def}
 X_1(\tau) \equiv  \eta \epsilon a^2  \zeta_p^*(\tau_0)  \zeta_p(\tau) \zeta'_q(\tau)^2 \, ,
 \end{equation} 
\begin{equation}
 \delta(\tau)  \equiv 2 \epsilon \eta a^2 \zeta'_q(\tau)^2
 \mathrm{Im}  \big[  \zeta_p^*(\tau_0)  \zeta_p(\tau)   \big]  \, ,
\end{equation}
and
\begin{equation}
 \beta (\tau) \equiv 2 \epsilon \eta a^2 \zeta'_q(\tau) \zeta_q(\tau)
 \mathrm{Im}
 \big[  \zeta_p^*(\tau_0)  \zeta'_p(\tau)   \big] \, .
 \end{equation}
 Note that there are additional subleading terms containing $\zeta'_p(\tau_2)$ which are not included in Eq. (\ref{loop-cc}). This is because these terms are suppressed by a factor $p^2$ compared to term denoted by $X_1(\tau_2)$.\footnote{In the analysis of \cite{Firouzjahi:2023aum}, these subleading terms are denoted by $X_2$, see discussions after Eq. (A.10) in \cite{Firouzjahi:2023aum}.  }

Looking at the expressions of $\delta (\tau)$ and $\beta(\tau)$, we note that  
$\delta (\tau)$ is originated from the commutator (\ref{com2b}) while $\beta(\tau)$
is originated from the commutator (\ref{com1}). Consequently, in the analysis of  \cite{Fumagalli:2023hpa} who uses Eq. (\ref{com2}) instead of (\ref{com2b}), 
the term containing $\delta(\tau)$ does not exist.
This corresponds to setting $c_1=0$. However, in our analysis we have $c_1=1$.
We have checked that if we set $c_1=0$, then Eq. (\ref{loop-cc}) agrees exactly with the corresponding result in \cite{Fumagalli:2023hpa} (Eq. (50) in \cite{Fumagalli:2023hpa}). 

Proceeding similarly, we obtain 
\ba
\label{loop-dc}
\big \langle \zeta_\bfp  \zeta_{\bfp'}  \big\rangle'_{[d,c]} = 
 8  M_P^4  \int_{\tau_s}^{\tau_e} d \tau_1 \int_{\tau_s}^{\tau_1} d \tau_2  \int \frac{d^3 \bfq}{(2 \pi)^3} q^2 \mathrm{Im} \Big[ Y^* ( \tau_2) \Big(c_1 \, \delta(\tau_1) + 2  \beta(\tau_1)   \Big) \Big]  \, ,
\ea
in which 
 \begin{equation}
\label{Y-def}
 Y(\tau) \equiv  \eta \epsilon a^2  
 \zeta_p^*(\tau_0)  \zeta_p(\tau) \zeta_q(\tau)^2 \, .
 \end{equation} 
As in previous case, if we set $c_1=0$, the above result agrees exactly with 
$\big \langle \zeta_\bfp  \zeta_{\bfp'}  \big\rangle_{[d,c]}$ obtained in  \cite{Fumagalli:2023hpa}. 
 
Finally, calculating $\big \langle \zeta_\bfp  \zeta_{\bfp'}  \big\rangle_{[c,d]}$, we obtain
\ba
\label{loop-cd}
\big \langle \zeta_\bfp  \zeta_{\bfp'}  \big\rangle'_{[c,d]} = 
 -16  M_P^4  \int_{\tau_s}^{\tau_e} d \tau_1 \int_{\tau_s}^{\tau_1} d \tau_2 \epsilon(\tau_1) a(\tau_1)^2 
  \int \frac{d^3 \bfq}{(2 \pi)^3} q^2 \,   \mathrm{Im}  \big[  \zeta_p^*(\tau_0)  \zeta_p(\tau_1)   \big]
  \mathrm{Im} \big[ X_1^* ( \tau_2)   {\zeta_q}(\tau_1)^2 \big].  \nonumber 
\ea 

Combining the results for $\big \langle \zeta_\bfp  \zeta_{\bfp'}  \big\rangle_{[c,c]}$, 
$\big \langle \zeta_\bfp  \zeta_{\bfp'}  \big\rangle_{[d,c]}$ and $\big \langle \zeta_\bfp  \zeta_{\bfp'}  \big\rangle_{[c,d]}$,  the total one-loop correction at the cubic order is obtained to be 
\ba
\label{loop-total-a}
\big \langle \zeta_\bfp  \zeta_{\bfp'}  \big\rangle'_{\bf{H_3}} = 
 8  M_P^4  \int_{\tau_s}^{\tau_e} d \tau_1 \int_{\tau_s}^{\tau_1} d \tau_2  \int \frac{d^3 \bfq}{(2 \pi)^3}  {\cal F}(\tau_1, \tau_2; q)    \, ,
\ea
in which 
\ba
\label{F-def}
{\cal F}(\tau_1, \tau_2; q)  \equiv 
 \mathrm{Im}  \left \{ X_1^*( \tau_2) \Big[  \Big(c_1 \, \delta(\tau_1) + 2  \beta(\tau_1) \Big)
 \big( 1- f_q^*(\tau_2) \big) - c_1 f_q( \tau_1) \delta (\tau_1)  \Big]  \right \}  \, ,
\ea
and,
\ba
f_q(\tau) \equiv  q^2 \frac{Y}{X_1} =  \frac{q^2 \zeta_q^2}{{\zeta'_q}^2} \, .
\ea
As mentioned before, our result in Eq. (\ref{loop-total}) reduces to the result 
of \cite{Fumagalli:2023hpa} if we set $c_1=0$. In addition, performing the nested integral, one can show that the last term in ${\cal F}$, containing $f_q( \tau_1) \delta (\tau_1)$, is subleading which  agrees with the conclusion 
in \cite{Fumagalli:2023hpa} that the 
contribution of $\big \langle \zeta_\bfp  \zeta_{\bfp'}  \big\rangle_{[c,d]}$ is subleading compared to $\big \langle \zeta_\bfp  \zeta_{\bfp'}  \big\rangle_{[c,c]}$ and   $\big \langle \zeta_\bfp  \zeta_{\bfp'}  \big\rangle_{[d,c]}$.

Finally,  Eq. (\ref{loop-total-a}) agrees with our earlier  result in \cite{Firouzjahi:2023aum} which was obtained using a somewhat different method to implement the 
in-in analysis. More specifically, in \cite{Firouzjahi:2023aum} the in-in analysis 
 is performed  as follows  
\ba
   \langle \zeta_{\bfp}(\tau_0) \zeta_{\bfp'}(\tau_0) \rangle_{\Ha} =    \langle \zeta_{\bfp}(\tau_0) \zeta_{\bfp'}(\tau_0) \rangle_{(2,0)} +   \langle \zeta_{\bfp}(\tau_0) \zeta_{\bfp'}(\tau_0) \rangle_{(1,1)} +    \langle \zeta_{\bfp}(\tau_0) \zeta_{\bfp'}(\tau_0) \rangle_{(0, 2)}
  \ea
  in which 
\ba
  \label{20-int}
\langle \zeta_{\bfp}(\tau_0) \zeta_{\bfp'}(\tau_0) \rangle_{(2,0)} &=&
- \int_{-\infty}^{\tau_0} d \tau_1 \int_{-\infty}^{\tau_1} d \tau_2
\big \langle \Ha (\tau_2)  \Ha (\tau_1) \zeta_{\bfp}(\tau_0) \zeta_{\bfp'}(\tau_0) 
 \big \rangle   \nonumber\\
&=& \langle \zeta_{\bfp}(\tau_0) \zeta_{\bfp'}(\tau_0) 
\rangle^\dagger_{(0,2)}\, ,
\ea
and 
 \ba
  \label{11-int}
\langle \zeta_{\bfp}(\tau_0) \zeta_{\bfp'}(\tau_0) \rangle_{(1,1)} =
 \int_{-\infty}^{\tau_0} d \tau_1 \int_{-\infty}^{\tau_0} d \tau_2
\big \langle \Ha (\tau_1)   \zeta_{\bfp}(\tau_0) \zeta_{\bfp'}(\tau_0) 
\Ha (\tau_2) \big \rangle   \, .
\ea
Combining all contributions one obtains the same result as Eq. (\ref{loop-total-a}), see  \cite{Firouzjahi:2023aum} for detail derivations.

\subsection{Loop Cancellation?  } 
\label{cancellation}

Our goal here is to examine the loop cancellation at the cubic order as advocated in \cite{Fumagalli:2023hpa}.  Plugging the mode functions (\ref{mode2}) and (\ref{mode3}) in Eq. (\ref{loop-total-a}) and performing the nested integrals for the range $\tau_s \leq \tau_2 \leq \tau_1 \leq \tau_e$ and 
$-\frac{1}{\tau_s} \leq q \leq -\frac{1}{\tau_e}$, 
we obtain 
\ba
\label{loop-total}
\big \langle \zeta_\bfp  \zeta_{\bfp'}  \big\rangle'_{\bf{H_3}} = 
\frac{9 c_1(h-12) }{8h} \big( \Delta N e^{6 \Delta N} \big)  \frac{H^4  }{  2 \pi^2 M_P^4 \epsilon_i^2 p^3 }   \, .
\ea
Correspondingly, the correction in one-loop power spectrum  from the cubic Hamiltonian $\Delta \calP_{\bf{H_3}}$ is obtained to be
\ba
\label{Delta P}
\Delta \calP_{\bf{H_3}} \equiv \frac{p^3}{2 \pi^2} \big \langle \zeta_\bfp^2    \big\rangle_{\bf{H_3}} = \frac{18 c_1 (h-12)}{h} \big( \Delta N e^{6 \Delta N} \big) 
\calP_{\mathrm{CMB}}^2 \, .
\ea
As expected, we see that if $c_1=0$, then the loop correction cancels to leading order as advocated in \cite{Fumagalli:2023hpa}. However, the consistent analysis requires $c_1=1$ and there is no loop cancellation. 

The above result is qualitatively consistent with the result obtained  \cite{Kristiano:2022maq}. However, with $h=-6$ which is the case studied in \cite{Kristiano:2022maq},   the above result is larger than the result obtained in \cite{Kristiano:2022maq} by a factor of 6. There may be a number of reasons for this numerical discrepancy. Note that in \cite{Kristiano:2022maq} they used the new variables $\zeta_n$ defined in \cite{Maldacena:2002vr} with $\zeta= \zeta_n + {\cal O} (\zeta_n^2)$ while here we work with $\zeta$. It is possible that the non-linear relation between $\zeta$ and $\zeta_n$ induces quartic interactions from the starting cubic interactions which were not taken into account in \cite{Kristiano:2022maq}. In addition,  the boundary term was not included in the analysis 
of \cite{Kristiano:2022maq} which may also contribute into the numerical mismatch. 

The conclusion is that there is no cancelation in one-loop correction at the  cubic order. The source of the disagreement with the conclusion of 
\cite{Fumagalli:2023hpa} is that $[ \zeta_\bfp(\tau_0), \zeta_\bfq(\tau)] \neq 0$ as summarized in  Eq. (\ref{com2b}). This is because one can not neglect the roles of the would-be decaying mode  which grows exponentially during the USR phase. 

Now we can compare our result Eq. (\ref{loop-total}) with the corresponding result obtained in \cite{Firouzjahi:2023aum}. We see that our Eq. (\ref{loop-total}) agrees exactly with the result obtained in  \cite{Firouzjahi:2023aum} when the integration\footnote{In performing the nested integrals,  two different strategies were considered in \cite{Firouzjahi:2023aum}. 
In the first strategy, one  calculates the  nested integral considering only the modes which become superhorizon during the USR phase. This means to cut the time integral in the range $-\frac{1}{q} \leq \tau_2 \leq \tau_1 \leq \tau_e $ so the lower bound of the integral is $-\frac{1}{q}$ instead of $\tau_s$. 
The second strategy is similar to what followed here (as in  \cite{Fumagalli:2023hpa}), integrating over all modes,  whether subhorizon or superhorizon during the USR phase, corresponding to  $\tau_s \leq \tau_2 \leq \tau_1 \leq \tau_e$. This corresponds to simply setting the lower bound of the time integral to be $\tau_s$ as 
given in Eq. (\ref{loop-total-a}). } is 
over $\tau_s \leq \tau_2 \leq \tau_1 \leq \tau_e$, see  Eq. (5.39) in \cite{Firouzjahi:2023aum}. This is not surprising, since the starting cubic Hamiltonian in both \cite{Firouzjahi:2023aum} and \cite{Fumagalli:2023hpa} is the same, as given in Eq. (\ref{H3-2}).

Finally, we comment that the contribution of the quartic Hamiltonian in loop correction  was calculated in  \cite{Firouzjahi:2023aum} which we present here: 
\begin{equation}
\label{quartic-power}
  \langle \zeta_{\bfp}^2 \rangle'_{\Hb} =  \frac{3}{8h} \big(  h^2 + 6h + 36
 \big) \big( {\Delta N}{e^{6 \Delta N}}  \big) \frac{H^4  }{ 2 \pi^2 M_P^4 \epsilon_i^2 p^3 } \, .
\end{equation}
We see that it has a somewhat different dependence on the sharpness parameter $h$ such that the quartic one-loop corrections scales linearly with $h$ for $|h| \gg1$. 

Combining the cubic and quartic one-loop corrections from Eqs. (\ref{quartic-power}) and (\ref{loop-total}), and setting $c_1=1$, 
the total one-loop correction is given by
\ba
\label{Delta P-total}
\Delta \calP_{\bf{H_3}+ \bf{H_4}} = (6h + 54) \big( \Delta N e^{6 \Delta N} \big) 
\calP_{\mathrm{CMB}}^2 \, .
\ea
We see that  the total one-loop correction scales linearly with $h$. There is no cancellation in total one-loop correction for a general value of $h$ except
at $h=-9$. Of course, there are subleading terms in one-loop contributions which were not included in our analysis here so we believe that even for $h=-9$, the one-loop cancellation does not occur.  Finally, for $h=-6$ which is the case  studied in \cite{Kristiano:2022maq, Kristiano:2023scm},  the total one-loop correction in Eq. (\ref{Delta P-total}) is larger than  the result in \cite{Kristiano:2022maq, Kristiano:2023scm} by a factor of 2.  Having said this, it is interesting that the final result, once the effects of the boundary terms and the quartic interaction are incorporated, is qualitatively in agreement with the results  of \cite{Kristiano:2022maq, Kristiano:2023scm}. 

As argued in \cite{Kristiano:2022maq, Kristiano:2023scm} the  loop corrections 
in the form of Eq. (\ref{Delta P-total}) can get out of control if one enhances the short scale power spectrum during the USR phase by a factor $10^7$ to generate the desired PBHs abundance. Furthermore, as argued in \cite{Firouzjahi:2023aum},  this gets even worse if one considers extreme sharp transitions  with $h \rightarrow -\infty$. However, for a mild transition with $h \sim \eta_V$,  the loop corrections will be slow-roll suppressed and the model is reliable for PBHs 
formation \cite{Firouzjahi:2023ahg, Riotto:2023hoz}.

\section{Summary and Discussions } 
\label{Summary}

In this work we have revisited the question of one-loop corrections in the setup which contains an intermediate phase of USR inflation. First, we have provided physical arguments on the reality of loop effects. More specifically, one may worry that large loop corrections on long modes induced from small scales  may be in conflict with the  notion of the decoupling of scales. We have tried to clarify this puzzle. We have argued that the non-linear couplings between the long and short modes generate a second order source term for the evolution of the long mode perturbations. On the other hand, the long mode rescales the background coordinate so its effects can be viewed as a modulation of the short mode power spectrum. These two effects combine to induce a non-trivial back-reaction on the long mode itself which can be viewed as the source of the  loop corrections \cite{Riotto:2023gpm}. In order for the loop corrections to be noticeable we require a significant scale-dependence for the power spectrum of the short modes. This is guaranteed in the USR phase as the power spectrum of the modes which leave the horizon during the USR phase  experiences a rapid rise like $\calP_\zeta \propto q^4$.

In the first part of this work we have found that our expression for the loop correction Eq. (\ref{loop}) has the same structure as advocated in   \cite{Tada:2023rgp}. However, we disagree with the argument in \cite{Tada:2023rgp} that the contribution of $q_{\mathrm{max}}$ is negligible after performing the $i \varepsilon$ prescription. Indeed, a natural prescription for $q_{\mathrm{max}}$ is $q_{\mathrm{max}}= q_e$ as advocated in \cite{Kristiano:2022maq, Kristiano:2023scm}.  This leads to the expected result 
Eq. (\ref{loop2}). However, if we follow the prescription of \cite{Tada:2023rgp} and push $q_{\mathrm{max}}$ to the maximum allowed value, then  the power spectrum has a quadratic divergence in the UV region with rapid small oscillations superimposed on top of it. We argued that these oscillations are harmless as they are much smaller than the overall quadratic divergence. Indeed, the situation here is similar to the standard QFTs in which one has to employ a renormalization scheme, such as the  dimensional regularization approach, to regularize and 
renormalize the divergent power spectrum. 
For this to be consistent, one requires the loop corrections to be perturbatively under control order by order. Consequently,  we need the fractional one-loop correction in Eq. (\ref{loop2}) with the amplitude $e^{6 \Delta N} \calP_{\mathrm{CMB}}$ to be small.

In the second part of this work we have revisited the claim in \cite{Fumagalli:2023hpa}  that the loop contributions cancel out to leading order when using the cubic Hamiltonian.  We have highlighted the important roles played by the would-be decaying mode during the USR phase. 
As it is well known, the decaying mode grows exponentially during the USR period which is the main reason behind the violation of  Maldacena's consistency condition \cite{Namjoo:2012aa}. Correspondingly, one can not simply take it for granted that $[ \zeta_\bfp(\tau_0), \zeta_\bfq(\tau)] \neq 0$ so one should use  Eq. (\ref{com2b}) instead of (\ref{com2}). The contribution of Eq. (\ref{com2b}) in our analysis is captured by the term $\delta$ in Eq. (\ref{F-def}). To follow the contribution of  the term $\delta$ we have inserted the fiducial parameter $c_1$ in the follow up analysis. We have checked that if $c_1=0$, then one reproduces the result of \cite{Fumagalli:2023hpa} in which the loop corrections  by cubic interactions cancel out to leading order. However, in the correct treatment with 
 $c_1=1$, the loop correction does not cancel out as seen explicitly in Eq. (\ref{loop-total}).

As the  one-loop corrections are genuine and are 
not canceled out  then one has to worry about 
their cosmological implications. In particular, it may not be easy to generate PBHs in the models employing an intermediate  phase of USR inflation as highlighted in \cite{Kristiano:2022maq, Kristiano:2023scm}. The amplitude of loop corrections scales linearly with the sharpness parameter $h$. Correspondingly,  for sharp transitions the loop corrections can get out of control for $\Delta N >1$. However, as shown in  \cite{Firouzjahi:2023ahg}, the loop corrections will be slow-roll suppressed if the transition is mild. Another interesting question is the loop effects on the bispectrum. The 
experience with the case of power spectrum suggests that  the loop corrections can have significant impacts in $f_{NL}$ parameter on large CMB scales as well. This is a non-trivial question since the corresponding in-in analysis involves higher order nested  integrals. Another question of interest is to look at two-loops and higher orders loops corrections for both power spectrum and bispectrum.  We would like to come back to these questions in future.

\vspace{0.9cm}
   
 {\bf Acknowledgments:}  We thank   Antonio Riotto, Mohammad Hossein Namjoo, Jacopo Fumagalli  and Sina Hooshangi for helpful discussions and correspondences.   We are grateful to Jason Kristiano for many insightful comments and discussions. This work is supported by INSF  of Iran under the grant  number 4025208.

 \vspace{0.5cm}
  


{}


\begin{thebibliography}{}




\bibitem{Kristiano:2022maq}
J.~Kristiano and J.~Yokoyama,
[arXiv:2211.03395 [hep-th]].

\bibitem{Kristiano:2023scm}
J.~Kristiano and J.~Yokoyama,
[arXiv:2303.00341 [hep-th]].


\bibitem{Riotto:2023hoz}
A.~Riotto,
[arXiv:2301.00599 [astro-ph.CO]].

\bibitem{Riotto:2023gpm}
A.~Riotto,
[arXiv:2303.01727 [astro-ph.CO]].

\bibitem{Choudhury:2023vuj}
S.~Choudhury, M.~R.~Gangopadhyay and M.~Sami,
[arXiv:2301.10000 [astro-ph.CO]].

\bibitem{Choudhury:2023jlt}
S.~Choudhury, S.~Panda and M.~Sami,
[arXiv:2302.05655 [astro-ph.CO]].


\bibitem{Choudhury:2023rks}
S.~Choudhury, S.~Panda and M.~Sami,
[arXiv:2303.06066 [astro-ph.CO]].

\bibitem{Choudhury:2023hvf}
S.~Choudhury, S.~Panda and M.~Sami,
JCAP \textbf{08}, 078 (2023), 
[arXiv:2304.04065 [astro-ph.CO]].

\bibitem{Firouzjahi:2023aum}
H.~Firouzjahi,
JCAP \textbf{10}, 006 (2023), 
[arXiv:2303.12025 [astro-ph.CO]].
  
\bibitem{Motohashi:2023syh}
H.~Motohashi and Y.~Tada,
[arXiv:2303.16035 [astro-ph.CO]].

\bibitem{Firouzjahi:2023ahg}
H.~Firouzjahi and A.~Riotto,
[arXiv:2304.07801 [astro-ph.CO]].

\bibitem{Tasinato:2023ukp}
G.~Tasinato,
Phys. Rev. D \textbf{108}, no.4, 043526 (2023), 
[arXiv:2305.11568 [hep-th]].

\bibitem{Franciolini:2023lgy}
G.~Franciolini, A.~Iovino, Junior., M.~Taoso and A.~Urbano,
[arXiv:2305.03491 [astro-ph.CO]].

\bibitem{Firouzjahi:2023btw}
H.~Firouzjahi,
Phys. Rev. D \textbf{108}, no.4, 043532 (2023), 
[arXiv:2305.01527 [astro-ph.CO]].

\bibitem{Maity:2023qzw}
S.~Maity, H.~V.~Ragavendra, S.~K.~Sethi and L.~Sriramkumar,
[arXiv:2307.13636 [astro-ph.CO]].

\bibitem{Cheng:2023ikq}
S.~L.~Cheng, D.~S.~Lee and K.~W.~Ng,
[arXiv:2305.16810 [astro-ph.CO]].


\bibitem{Fumagalli:2023loc}
J.~Fumagalli, S.~Bhattacharya, M.~Peloso, S.~Renaux-Petel and L.~T.~Witkowski,
[arXiv:2307.08358 [astro-ph.CO]].

\bibitem{Nassiri-Rad:2023asg}
A.~Nassiri-Rad and K.~Asadi,
[arXiv:2310.11427 [astro-ph.CO]].

\bibitem{Meng:2022ixx}
D.~S.~Meng, C.~Yuan and Q.~g.~Huang,
Phys. Rev. D \textbf{106}, no.6, 063508 (2022), 
[arXiv:2207.07668 [astro-ph.CO]].




\bibitem{Cheng:2021lif}
S.~L.~Cheng, D.~S.~Lee and K.~W.~Ng,
Phys. Lett. B \textbf{827}, 136956 (2022), 
[arXiv:2106.09275 [astro-ph.CO]].

\bibitem{Ivanov:1994pa}
P.~Ivanov, P.~Naselsky and I.~Novikov,
Phys. Rev. D \textbf{50}, 7173-7178 (1994). 

\bibitem{Garcia-Bellido:2017mdw}
J.~Garcia-Bellido and E.~Ruiz Morales,
Phys. Dark Univ. \textbf{18}, 47-54 (2017), 
[arXiv:1702.03901 [astro-ph.CO]].

\bibitem{Biagetti:2018pjj}
M.~Biagetti, G.~Franciolini, A.~Kehagias and A.~Riotto,
JCAP \textbf{07}, 032 (2018). 

\bibitem{Khlopov:2008qy}
M.~Y.~Khlopov,
Res. Astron. Astrophys. \textbf{10}, 495-528 (2010), 
[arXiv:0801.0116 [astro-ph]].

\bibitem{Ozsoy:2023ryl}
O.~\"Ozsoy and G.~Tasinato,
Universe \textbf{9}, no.5, 203 (2023),
[arXiv:2301.03600 [astro-ph.CO]].


\bibitem{Byrnes:2021jka}
C.~T.~Byrnes and P.~S.~Cole,
[arXiv:2112.05716 [astro-ph.CO]].


\bibitem{Fumagalli:2023hpa}
J.~Fumagalli,
[arXiv:2305.19263 [astro-ph.CO]].

\bibitem{Tada:2023rgp}
Y.~Tada, T.~Terada and J.~Tokuda,
[arXiv:2308.04732 [hep-th]].


\bibitem{Maldacena:2002vr} 
  J.~M.~Maldacena,
  JHEP {\bf 0305}, 013 (2003),
  [astro-ph/0210603].

\bibitem{Kinney:2005vj} 
  W.~H.~Kinney,
  Phys.\ Rev.\ D {\bf 72}, 023515 (2005),
  [gr-qc/0503017].
  
\bibitem{Morse:2018kda}
M.~J.~P.~Morse and W.~H.~Kinney,
Phys. Rev. D \textbf{97}, no.12, 123519 (2018), 
[arXiv:1804.01927 [astro-ph.CO]].

\bibitem{Lin:2019fcz}
W.~C.~Lin, M.~J.~P.~Morse and W.~H.~Kinney,
JCAP \textbf{09}, 063 (2019), 
[arXiv:1904.06289 [astro-ph.CO]].


\bibitem{Creminelli:2004yq}
P.~Creminelli and M.~Zaldarriaga,
JCAP \textbf{10}, 006 (2004), 
[arXiv:astro-ph/0407059 [astro-ph]].

\bibitem{Namjoo:2012aa} 
  M.~H.~Namjoo, H.~Firouzjahi and M.~Sasaki,
  Europhys.\ Lett.\  {\bf 101}, 39001 (2013),
  [arXiv:1210.3692 [astro-ph.CO]].

\bibitem{Martin:2012pe}
J.~Martin, H.~Motohashi and T.~Suyama,
Phys. Rev. D \textbf{87}, no.2, 023514 (2013), 
[arXiv:1211.0083 [astro-ph.CO]].
  
  
\bibitem{Chen:2013aj} 
  X.~Chen, H.~Firouzjahi, M.~H.~Namjoo and M.~Sasaki,
  Europhys.\ Lett.\  {\bf 102}, 59001 (2013), 
  [arXiv:1301.5699 [hep-th]].
 
\bibitem{Chen:2013eea} 
  X.~Chen, H.~Firouzjahi, E.~Komatsu, M.~H.~Namjoo and M.~Sasaki,
  JCAP {\bf 1312}, 039 (2013),  [arXiv:1308.5341 [astro-ph.CO]].
  
\bibitem{Akhshik:2015nfa}
M.~Akhshik, H.~Firouzjahi and S.~Jazayeri,
JCAP \textbf{07}, 048 (2015), 
[arXiv:1501.01099 [hep-th]].
  
\bibitem{Akhshik:2015rwa}
M.~Akhshik, H.~Firouzjahi and S.~Jazayeri,
JCAP \textbf{12}, 027 (2015), 
[arXiv:1508.03293 [hep-th]].



\bibitem{Mooij:2015yka}
S.~Mooij and G.~A.~Palma,
JCAP \textbf{11}, 025 (2015), 
[arXiv:1502.03458 [astro-ph.CO]].

\bibitem{Bravo:2017wyw}
R.~Bravo, S.~Mooij, G.~A.~Palma and B.~Pradenas,
JCAP \textbf{05}, 024 (2018), 
[arXiv:1711.02680 [astro-ph.CO]].

\bibitem{Finelli:2017fml}
B.~Finelli, G.~Goon, E.~Pajer and L.~Santoni,
Phys. Rev. D \textbf{97}, no.6, 063531 (2018), 
[arXiv:1711.03737 [hep-th]].

\bibitem{Passaglia:2018ixg}
S.~Passaglia, W.~Hu and H.~Motohashi,
Phys. Rev. D \textbf{99}, no.4, 043536 (2019), 
[arXiv:1812.08243 [astro-ph.CO]].

\bibitem{Pi:2022ysn}
S.~Pi and M.~Sasaki,
Phys. Rev. Lett. \textbf{131}, no.1, 011002 (2023), 
[arXiv:2211.13932 [astro-ph.CO]].

\bibitem{Firouzjahi:2023xke}
H.~Firouzjahi and A.~Riotto,
[arXiv:2309.10536 [astro-ph.CO]].

\bibitem{Cai:2018dkf}
Y.~F.~Cai, X.~Chen, M.~H.~Namjoo, M.~Sasaki, D.~G.~Wang and Z.~Wang,
JCAP \textbf{05}, 012 (2018), 
[arXiv:1712.09998 [astro-ph.CO]].

\bibitem{Cai:2022erk}
Y.~F.~Cai, X.~H.~Ma, M.~Sasaki, D.~G.~Wang and Z.~Zhou,
JCAP \textbf{12}, 034 (2022), 
[arXiv:2207.11910 [astro-ph.CO]].

\bibitem{Arroja:2011yj}
F.~Arroja and T.~Tanaka,
JCAP \textbf{05}, 005 (2011), 
[arXiv:1103.1102 [astro-ph.CO]].



\bibitem{Byrnes:2018txb}
C.~T.~Byrnes, P.~S.~Cole and S.~P.~Patil,
JCAP \textbf{06}, 028 (2019),
[arXiv:1811.11158 [astro-ph.CO]].

\bibitem{Cole:2022xqc}
P.~S.~Cole, A.~D.~Gow, C.~T.~Byrnes and S.~P.~Patil,
[arXiv:2204.07573 [astro-ph.CO]].

\bibitem{Carrilho:2019oqg}
P.~Carrilho, K.~A.~Malik and D.~J.~Mulryne,
Phys. Rev. D \textbf{100}, no.10, 103529 (2019),
[arXiv:1907.05237 [astro-ph.CO]].

\bibitem{Ozsoy:2021pws}
O.~\"Ozsoy and G.~Tasinato,
Phys. Rev. D \textbf{105}, no.2, 023524 (2022),
[arXiv:2111.02432 [astro-ph.CO]].

\bibitem{Pi:2022zxs}
S.~Pi and J.~Wang,
JCAP \textbf{06}, 018 (2023),
[arXiv:2209.14183 [astro-ph.CO]].


\bibitem{Weinberg:2005vy}
S.~Weinberg,
Phys. Rev. D \textbf{72}, 043514 (2005)\, .
[arXiv:hep-th/0506236 [hep-th]].

\bibitem{Senatore:2009cf}
L.~Senatore and M.~Zaldarriaga,
JHEP \textbf{12}, 008 (2010), 
[arXiv:0912.2734 [hep-th]].

\bibitem{Pimentel:2012tw}
G.~L.~Pimentel, L.~Senatore and M.~Zaldarriaga,
JHEP \textbf{07}, 166 (2012), 
[arXiv:1203.6651 [hep-th]].












\end{thebibliography}
\end{document}